\documentclass[reprint,amsmath,amssymb,aps,prc]{revtex4-2}

\usepackage{graphicx}
\usepackage{dcolumn}
\usepackage{bm}
\usepackage{epstopdf}
\usepackage[mathlines]{lineno}
\usepackage{harpoon}
\usepackage{makecell}
\usepackage{subfigure}
\usepackage{hyperref}
\usepackage{color}
\usepackage{braket}
\usepackage{amsmath}
\usepackage{mathrsfs}
\usepackage{appendix}
\usepackage{multirow}
\usepackage{float}
\usepackage{enumerate}
\usepackage{ulem}
\usepackage{slashed}

\begin{document}

\title{The odd-even differences in stability peninsula for $106 \leqslant Z \leqslant 112$ region with the deformed relativistic Hartree-Bogoliubov theory in continuum}
    \author{Xiao-Tao He}
	\email{hext@nuaa.edu.cn}
  	\affiliation{College of Materials Science and Technology, Nanjing University of Aeronautics and Astronautics, Nanjing 210016, China}	
 	\author{Jia-Wei Wu}
        \email{wjw1998@nuaa.edu.cn}
 	\affiliation{College of Materials Science and Technology, Nanjing University of Aeronautics and Astronautics, Nanjing 210016, China}
	\author{Kai-Yuan Zhang}
	\email{zhangky@caep.cn}
    \affiliation{Institute of Nuclear Physics and Chemistry, China Academy of Engineering Physics, Mianyang, Sichuan 621900, China}
     	\author{Cai-Wan Shen}
	\email{cwshen@zjhu.edu.cn}
 	\affiliation{School of Science, Huzhou University, Huzhou 313000, China}
	
	\date{\today}
	
\begin{abstract}
The predictive power of the deformed relativistic Hartree-Bogoliubov theory in continuum (DRHBc) with density functional PC-PK1 is demonstrated for superheavy region ($101 \leqslant Z \leqslant 120$) by comparing with available experimental and evaluated data in the AME2020.
The DRHBc theory predicts 93 bound nuclei beyond the drip line $N = 258$ in the region of $106 \leqslant Z \leqslant 112$, which form a stability peninsula.
The odd-even differences between odd-$N$ and even-$N$ nuclei are remarkable in the stability peninsula;
the number of bound odd-$N$ nuclei is less than that of bound even-$N$ nuclei, and the one-neutron separation energy of an odd-$N$ nucleus is smaller than those of its neighboring even-$N$ nuclei due to the blocking effect.
The deformation effect is indispensable for the reentrant stability beyond the drip line by significantly affecting the structure of single-particle levels around the Fermi energy.
The interplay between deformation and pairing effects affects the position where the odd-$N$ nucleus becomes bound in the stability peninsula.
By examining the deformation effect at different orders, it is found that quadrupole deformation makes leading contribution to the appearance of stability peninsula and the effects of hexadecapole and hexacontatetrapole deformations are nonnegligible.
\end{abstract}

\maketitle

\section{Introduction}{\label{Sec:introduction}}

Exploration of the limit of the nuclear existence is a priority topic in nuclear physics \cite{Thoennessen2013RPP,Thoennessen2016,Erler2012Nature,Xia2018ADNDT}.
Along or near the $\beta$-stability line, the stable nuclei form the valley of stability.
Starting from the $\beta$-stability line, the nuclear binding energy keeps increasing with neutron adding until the neutron drip line, where the binding energy begins to fall.
There is no enough binding energy to prevent the last neutron(s) from escaping the nucleus at the neutron drip line, which depicts the boundary of the nuclear landscape on the neutron-rich side.
However, it is pointed out in Ref. \cite{Stoitsov2003PRC} that the binding energy may increase again with a further increase of the neutron number beyond the neutron drip line.
This phenomenon of reentrant stability against particle emission might lead to a novel \textit{peninsula} adjacent to the nuclear landscape \cite{Stoitsov2003PRC}.

To date, the existence of more than 3300 nuclides has been confirmed experimentally \cite{Thoennessen2023,DNP}, and the masses of about 2500 among them have been measured \cite{Wang2021CPC,Kondev2021CPC,Huang20212CPC}.
The proton-rich boundary of the nuclear territory has been determined up to neptunium ($Z=93$) \cite{Zhang2019PRL}, but the neutron-rich boundary is known only up to neon ($Z=10$) \cite{Ahn2019PRL}.
The experimental exploration of very neutron-rich nuclei is extremely challenging, and most of neutron-rich nuclei far from the valley of stability seem still beyond the experimental capability in the foreseeable future.
Thus, for heavier elements, the exploration of the limits of nuclear existence relies on theoretical models.

The phenomenon of reentrant stability beyond the neutron drip line has been predicted in several studies.
Due to the shell effects at neutron closure, the regions of reentrant stability around $Z = 60$, 70, and 100 have been found by Skyrme Hartree-Fock-Bogoliubov calculations in the transformed harmonic oscillator (THO) basis \cite{Erler2012Nature,Stoitsov2003PRC}.
Relativistic Hartree-Bogoliubov calculations in the harmonic oscillator (HO) basis also predicted similar phenomena, which were attributed to the local changes in the shell structure induced by deformation \cite{Afanasjev2013PLB}.
The THO basis \cite{Stoitsov1998PRC,Stoitsov1998PRC2} provides an improved asymptotic behavior compared with the HO basis that is not appropriate for the description of nuclei near the drip line \cite{Dobaczewski1996PRC,Meng1998NPA,Zhou2000CPL,Zhou2003PRC}.
However, both of them are less suitable than the coordinate-space solution for the description of systems with very diffuse density distributions \cite{Zhou2000CPL,Zhang2013PRC}.
The reentrant stability has also been studied with the Skyrme Hartree-Fock + BCS theory \cite{Gridnev2008IJMPE,Walter2010NPA,Tarasov2013IJMPE}, but the BCS theory is incapable of describing the pairing correlations in exotic nuclei \cite{Meng2006PPNP,Meng2015JPG}.
Instead, the Bogoliubov transformation provides a well-established method to treat in a unified way the pairing correlations in both stable and exotic nuclei \cite{Dobaczewski1984NPA}.

The covariant density functional theory (CDFT) has successfully described a variety of nuclear phenomena \cite{Ring1996PPNP,Vretenar2005PR,Meng2006PPNP,Niksic2011PPNP,Meng2013FP,Meng2015JPG,Zhou2016PhysScr}.
Based on the CDFT, properly considering pairing correlations and continuum effects, the relativistic continuum Hartree-Bogoliubov (RCHB) theory was developed with spherical symmetry assumed \cite{Meng1998NPA,Meng1996PRL}.
The RCHB theory has successfully described and predicted many phenomena \cite{Meng1996PRL,Meng1998PRL,Zhang2002CPL,Meng1998PRC,Meng1999PRC,Zhang2005NPA,Kuang2023EPJA,Wu2024PRC}.
Based on the RCHB theory, the first nuclear mass table including continuum effects was constructed, and the continuum effects on the limits of nuclear landscape were investigated \cite{Xia2018ADNDT}.

Except for doubly magic nuclei, most nuclei in the nuclear landscape deviate from spherical shape.
Simultaneously including the effects of deformation, pairing correlations, and continuum, the deformed relativistic Hartree-Bogoliubov theory in continuum (DRHBc) was developed \cite{Zhou2010PRC,Li2012PRC}.
The advantages of the DRHBc theory have been highlighted by prosperous applications, e.g., to deformed halo nuclei \cite{Sun2018PLB,Zhang2019PRC,Yang2021PRL,Sun2021PRC(1),Zhang2023PRC(L1),Zhang2023PLB,Zhang2023PRC(L2),An2024PLB}.
Based on the DRHBc theory, the nuclear mass table for even-even nuclei has been constructed \cite{Zhang2022ADNDT}, and the peninsula of stability beyond the two-neutron drip line has been predicted \cite{Zhang2021PRC,Pan2021PRC,He2021CPC}.
In particular, the predictive power of the DRHBc theory combined with density functional PC-PK1 \cite{Zhao2010PRC} for the masses of superheavy nuclei has been demonstrated \cite{Zhang2021PRC}.

In previous studies \cite{Zhang2021PRC,Pan2021PRC,He2021CPC}, the phenomenon of reentrant stability based on the DRHBc theory has been discussed for even-even nuclei.
It is unclear whether or not there are bound odd nuclei in the stability peninsula.
It is therefore necessary to investigate the masses of odd-mass and odd-odd nuclei. The blocking effect of the unpaired nucleon(s) should be considered \cite{Ring2004}, which may play an important role in nuclear ground-state properties \cite{Sunevaluated data in RefNPA,Nakada2018PRC,Kasuya2020PTEP}.
The DRHBc theory has been extended to incorporate the blocking effect based on both meson-exchange \cite{Li2012CPL} and point-coupling \cite{Pan2022PRC} density functionals.

In this paper, we first examine the predictive ability of the DRHBc theory for even-even, odd-mass, and odd-odd nuclei in the superheavy region of $101 \leqslant Z \leqslant 120$. We then investigate the stability peninsula from Sg to Cn ($106 \leqslant Z \leqslant 112$), with the focus on exploring the odd-even differences between odd-$N$ and even-$N$ nuclei.
This paper is organized as follows.
In Sec. \ref{theoretical}, a brief theoretical framework is given.
Numerical details and the blocking procedure are introduced in Sec. \ref{numerical}.
Results and discussion are presented in Sec. \ref{results}.
Finally, a summary is provided in Sec. \ref{summary}.

\section{Theoretical framework}\label{theoretical}
The details of the DRHBc theory can be found in Refs. \cite{Li2012PRC,Pan2022PRC,Zhang2020PRC}.
Here we briefly present its formalism.
The relativistic Hartree-Bogoliubov (RHB) equation reads \cite{Kucharek1991ZPA}
\begin{align}
\left(\begin{array}{cc}
	h_D-\lambda_\tau & \Delta \\
	-\Delta^* & -h_D^*+\lambda_\tau
\end{array}\right)\left(\begin{array}{l}
	U_k \\
	V_k
\end{array}\right)=E_k\left(\begin{array}{c}
	U_k \\
	V_k
\end{array}\right),
 \end{align}
where $\lambda_\tau$ is the Fermi energy for neutrons or protons ($\tau = n$ or $p$), and $E_k$ and ($U_k$, $V_k$)$^T$ are the quasiparticle energy and wave function, respectively.
$h_D$ is the Dirac Hamiltonian,
\begin{align}
	h_D(\boldsymbol{r})=\alpha \cdot \boldsymbol{p}+V(\boldsymbol{r})+\beta[M+S(\boldsymbol{r})],
\end{align}
where $S(\boldsymbol{r})$ and $V(\boldsymbol{r})$ are the scalar and vector potentials, respectively.
$\Delta$ is the pairing potential,
\begin{align}
	\Delta\left(\boldsymbol{r}_1, \boldsymbol{r}_2\right)=V^{\mathrm{pp}}\left(\boldsymbol{r}_1, \boldsymbol{r}_2\right) \kappa\left(\boldsymbol{r}_1, \boldsymbol{r}_2\right),
\end{align}
with the pairing tensor $\kappa$ \cite{Ring2004} and a density-dependent pairing force of zero range,
\begin{align}
\resizebox{0.90\hsize}{!}{$V^{\mathrm{pp}}\left(\boldsymbol{r}_1, \boldsymbol{r}_2\right)=V_0 \frac{1}{2}\left(1-P^\sigma\right) \delta\left(\boldsymbol{r}_1-\boldsymbol{r}_2\right)\left(1-\frac{\rho\left(\boldsymbol{r}_1\right)}{\rho_{\mathrm{sat}}}\right),$}
\end{align}
in which $V_0$ is the pairing strength, $\rho_{\mathrm{sat}}$ denotes the saturation density of nuclear matter, and $\frac{1}{2}\left(1-P^\sigma\right)$ represents the projector for the spin $S = 0$ component in the pairing channel.
The pairing tensor, various densities, and potentials in coordinate space are expanded in terms of the Legendre polynomials,
\begin{align}
f(\boldsymbol{r})=\sum\limits_{\lambda} f_\lambda(r) P_\lambda(\cos \theta), \quad \lambda=0,2,4,\ldots
\label{Legendre}
\end{align}
To describe properly the possible large spatial extension of exotic nuclei, the RHB equations are solved in a Dirac Woods-Saxon basis \cite{Zhou2003PRC,Zhang2022PRC}, whose wave function has an appropriate asymptotic behavior.

For an odd-mass or odd-odd nucleus, one should take into account the blocking effect of the unpaired nucleon(s), which can be achieved in the RHB framework by the exchange of the quasiparticle wavefunctions $(V_{k_b}^*,U_{k_b}^*) \leftrightarrow (U_{k_b}, V_{k_b})$ and that of the energy $E_{k_b}\leftrightarrow -E_{k_b}$, in which $k_b$ stands for the blocked quasiparticle state. More details for the treatment of odd systems in the DRHBc theory can be found in Refs. \cite{Li2012CPL,Pan2022PRC}.

\section{NUMERICAL DETAILS}\label{numerical}

The calculations are carried out with the relativistic density functional PC-PK1 \cite{Zhao2010PRC}.
The pairing strength $V_0 = -325$ MeV fm$^3$ and the saturation density $\rho_{\mathrm{sat}} = 0.152$ fm$^{-3}$ in Eq. (4), and a pairing window of 100 MeV is adopted.
For the Dirac Woods-Saxon basis, the energy cutoff $E^+_{\mathrm{cut}}$ = 300 MeV and the angular momentum cutoff $J_{\mathrm{max}}$ = 23/2$\hbar$ are adopted.
In Eq. (5), the Legendre expansion is truncated at $\lambda_{\mathrm{max}} = 10$ \cite{Pan2019IJMPE}.
The above numerical details are the same as those used in the global DRHBc mass table calculations over the nuclear chart \cite{Zhang2022ADNDT,Pan2022PRC,Guo2024arXiv}.

In the present DRHBc calculations, the blocking effect is included via the equal filling approximation \cite{Perez-Martin2008PRC}, under which the time-odd components vanish and the time-reversal symmetry is conserved.
The orbital-fixed blocking and the automatic blocking have been implemented in the DRHBc theory \cite{Pan2022PRC}.
The calculation with orbital-fixed blocking procedure blocks separately the orbitals near the Fermi energy and takes the result with the lowest energy as the ground state \cite{Xia2018ADNDT}.
However, the computational cost of this procedure can be extremely high for deformed superheavy nuclei.
In the automatic one, the lowest quasiparticle orbital is always blocked during the iterative solution of the RHB equation, which can find in most cases the correct ground state with much less computational resource.
The DRHBc calculations are performed first with the automatic blocking and then, if necessary, with the orbital-fixed blocking.

\section{Results and discussion}\label{results}

\begin{figure}[htbp]
	\includegraphics[width=0.45\textwidth]{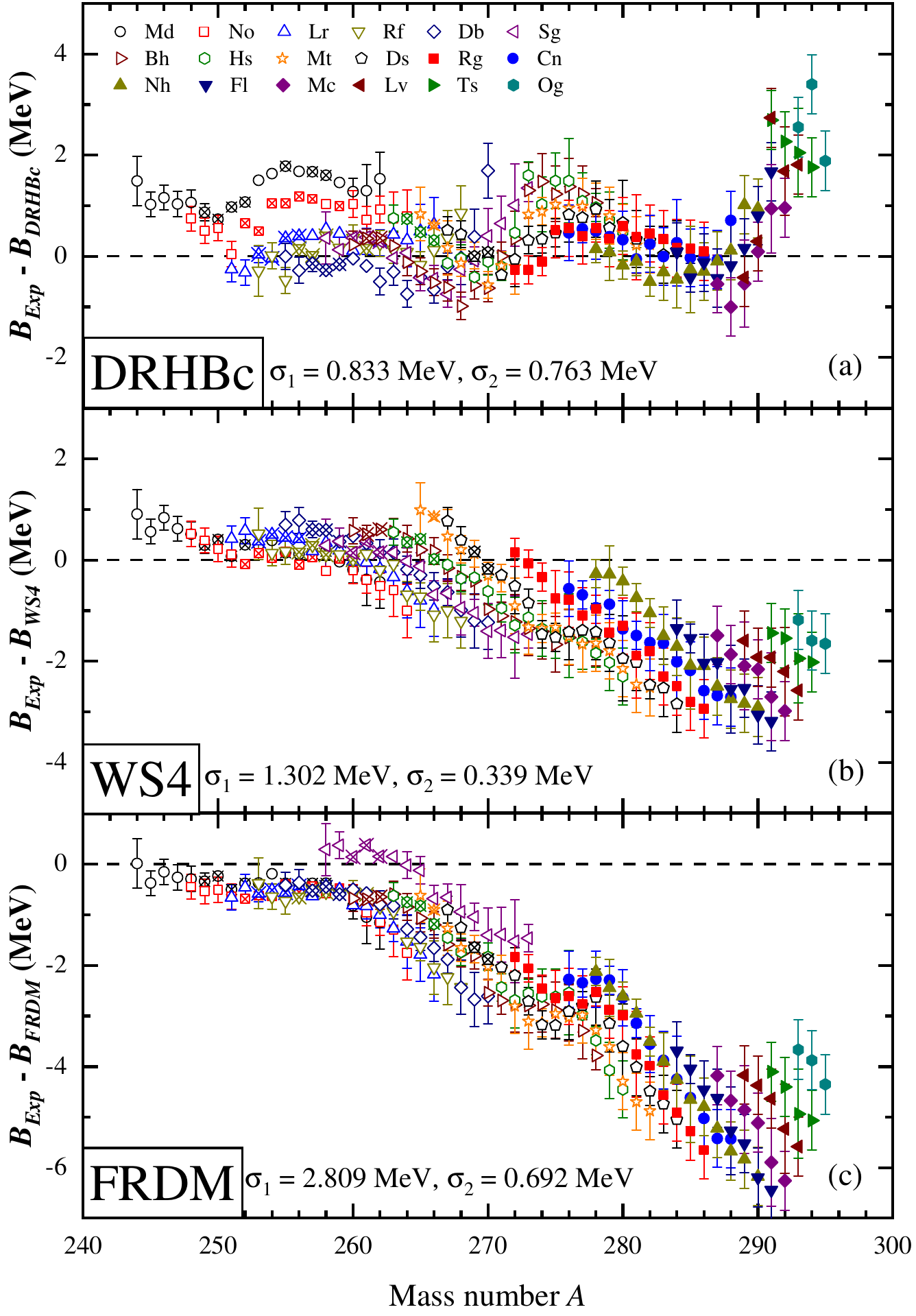}
	\caption{Binding energy difference between the AME2020 data and the calculations of (a) the DRHBc theory, (b) the WS4 mass model, and (c) the FRDM(2012) mass model for superheavy nuclei in the region of $101\leqslant Z\leqslant 118$. For $101\leqslant Z\leqslant 110$, nuclei with experimental and evaluated data in AME2020~\cite{Wang2021CPC} are shown by crossed and open symbols, respectively. For $111\leqslant Z\leqslant 118$, nuclei are shown by solid symbols with evaluated data in AME2020. $\sigma_1$ is the rms deviation from 240 available data in AME2020, and $\sigma_2$ is that from 36 experimental data.}
	\label{fig1}
\end{figure}

The predictive power of the DRHBc theory for even-even nuclei in the superheavy mass region has been demonstrated \cite{Zhang2021PRC} by comparing with the WS4 \cite{Wang2014PLB} and FRDM(2012) \cite{Moller2016ADNDT} mass models.
In this work, we further examine the DRHBc description for superheavy odd-mass and odd-odd nuclei.
Figure \ref{fig1}(a) shows the binding energy difference between the DRHBc calculated results and available data from AME2020 for nuclei with $101 \leqslant Z \leqslant 118$, in comparison with those for the popular macroscopic-microscopic mass models WS4 and FRDM in Figs. \ref{fig1}(b) and \ref{fig1}(c).
For 36 experimental data, the root-mean-square (rms) deviation $\sigma$ in the DRHBc calculations is 0.7663 MeV, larger than 0.339 MeV by WS4 and 0.692 MeV by FRDM.
When including 204 evaluated data in AME2020~\cite{Wang2021CPC}, the rms deviation of the total 240 data changes slightly to 0.833 MeV in the DRHBc calculations.
In contrast, the rms deviations for WS4 and FRDM increase to 1.302 and 2.809 MeV, respectively.

\begin{table*}
	\caption{The rms deviations (in MeV) of results based on the DRHBc theory, the WS4 mass model, and the FRDM(2012) mass model from the AME2020 data for total, even-even, odd-mass, and odd-odd nuclei in the region of $101 \leqslant Z \leqslant 118$. $\sigma_1$ and $\sigma_2$ represent the rms deviations from all available data in AME2020 and experimental data, respectively. The corresponding data numbers are given in parentheses.}
	\setlength\extrarowheight{5pt} 
	\setlength\arrayrulewidth{1pt} 
	\centering
	\begin{tabular}{lcccccccc}
		\hline
		&\multicolumn{2}{c}{All}
		&\multicolumn{2}{c}{Even-even}
		&\multicolumn{2}{c}{Odd-mass}
		&\multicolumn{2}{c}{Odd-odd}
		\\
		\hline
		&\multicolumn{1}{l}{~~~~$\sigma_1(240)$~~}
		&\multicolumn{1}{l}{~~$\sigma_2(36)$~~~~}
		&\multicolumn{1}{l}{~~~~$\sigma_1(57)$~~}
		&\multicolumn{1}{l}{~~$\sigma_2(10)$~~~~}
		&\multicolumn{1}{l}{~~~~$\sigma_1(118)$~~}
		&\multicolumn{1}{l}{~~$\sigma_2(18)$~~~~}
		&\multicolumn{1}{l}{~~~~$\sigma_1(65)$~~}
		&\multicolumn{1}{l}{~~$\sigma_2(8)$~~~~}
		\\
		DRHBc &~~0.833  &0.763~~  &~~0.777  &0.631~~  &~~0.848  &0.815~~  &~~0.883  &0.790~~
		\\
		WS4   &~~1.302  &0.339~~  &~~1.371  &0.159~~  &~~1.271  &0.308~~  &~~1.295  &0.522~~
		\\
		FRDM  &~~2.809  &0.692~~  &~~2.796  &0.864~~  &~~2.774  &0.648~~  &~~2.881  &0.526~~
		\\
		\hline
	\end{tabular}
	\label{table1}
\end{table*}

Table \ref{table1} shows the rms deviations of results based on the DRHBc theory, the WS4 mass model, and the FRDM mass model from the AME2020 data for all, even-even, odd-mass, and odd-odd nuclei in the region of $101 \leqslant Z \leqslant 118$.
It can be found that the description by the DRHBc theory is better than those by the WS4 and FRDM models  when including evaluated 204 data in AME2020.
In fact, the DRHBc description gradually becomes worse from even-even to odd-mass and then to odd-odd nuclei.
This is because in the present DRHBc calculations, the time-odd components, which might influence the rotational correction energy \cite{Zhao2012PRC}, are neglected.
It would be relevant to explore the effect of time-odd components on binding energies and rotational correction energies for odd nuclei by using the newly developed time-odd DRHBc theory with point-coupling density functionals \cite{Pan2024arXiv} in future works.
Furthermore, a better alternative to include the rotational correction is the two-dimensional collective Hamiltonian (2DCH), which can incorporate the beyond-mean-field dynamical correlations from rotations and vibrations for both well-deformed and near-spherical nuclei. The 2DCH has been implemented based on the DRHBc theory only for even-even nuclei \cite{Sun2022CPC,Zhang2023PRC}, and its future extension to odd nuclei is desirable.

\begin{figure*}[htbp]
	\includegraphics[width=1\textwidth]{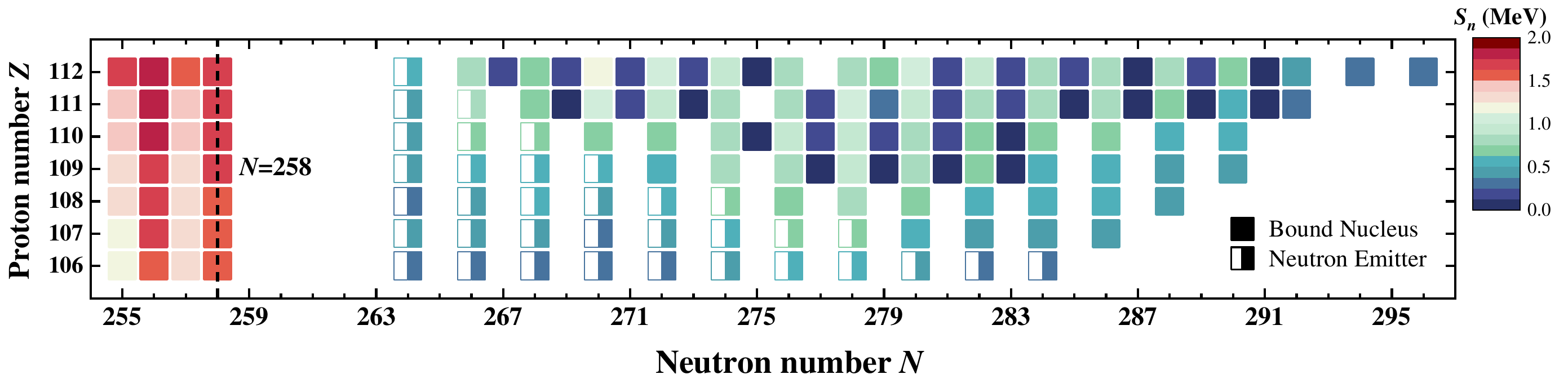}
	\caption{One-neutron separation energies for nuclei with $106 \leqslant Z \leqslant 112$ near the neutron drip line as a function of the neutron number $N$ and the proton number $Z$. The solid squares represent bound nuclei, and the semi-solid squares represent nuclei that are bound against one-neutron emission but unbound against multi-neutron emission. The dashed line depicts the predicted magic number $N = 258$.}
	\label{fig2}
\end{figure*}

The reliable predictive power of the DRHBc theory for superheavy nuclei enables the exploration of novel phenomena far from the stability valley.
In Refs. \cite{Zhang2021PRC,He2021CPC}, a stability peninsula consisting of even-even bound nuclei beyond the neutron drip line has been predicted in the region of $106 \leqslant Z\leqslant 112$.
It would be interesting to study whether odd-mass and odd-odd nuclei exist in this stability peninsula.
Figure \ref{fig2} illustrates the one-neutron separation energy $S_n$ for superheavy nuclei from Sg ($Z = 106$) to Cn ($Z = 112$) near the neutron drip line.
The sudden decrease of $S_n$ from positive to negative after $N = 258$ suggests it as a possible magic number.
This is consistent with the previous prediction made by other CDFT studies \cite{Zhang2005NPA,Afanasjev2013PLB,Li2014PLB}.
In Refs. \cite{Zhang2021PRC,He2021CPC}, the nucleus $^{370}$Sg$_{264}$ has a negative two-neutron separation energy, showing its instability against two-neutron emission.
Here, it is stable against one-neutron emission with a positive one-neutron separation energy.
The same situation also appears for other $N = 264$ isotones.
For even-$N$ nuclei in Bh, Hs, Mt, Ds, Rg, and Cn isotopic chains ($107\leqslant Z \leqslant 112$), the stability beyond the drip line can be found in the regions of $280 \leqslant N \leqslant 286$, $276 \leqslant N \leqslant 288$, $272 \leqslant N \leqslant 290$, $270 \leqslant N \leqslant 290$, $268 \leqslant N \leqslant 292$, and $266 \leqslant N \leqslant 296$, respectively.
However, the reentrant stability for odd-$N$ nuclei appears later and ends earlier than that for even-$N$ nuclei.
Specifically, there is no bound odd-$N$ nucleus with $Z=107$ and $108$.
Except two nuclei $^{386}$Rg$_{275}$ and $^{389}$Cn$_{277}$ unstable against one-neutron emission, the odd-$N$ nuclei in the regions of $277 \leqslant N \leqslant 283$, $275 \leqslant N \leqslant 283$, $269 \leqslant N \leqslant 291$, and $267 \leqslant N \leqslant 291$ are bound for $Z = 109, 110,111,$ and $112$, respectively.
The one-neutron separation energies of these bound odd-$N$ nuclei are obviously smaller than those of their even-$N$ neighbors, i.e., significant odd-even difference can be found in the one-neutron separation energy.

\begin{figure}[htbp]
	\includegraphics[scale=0.270]{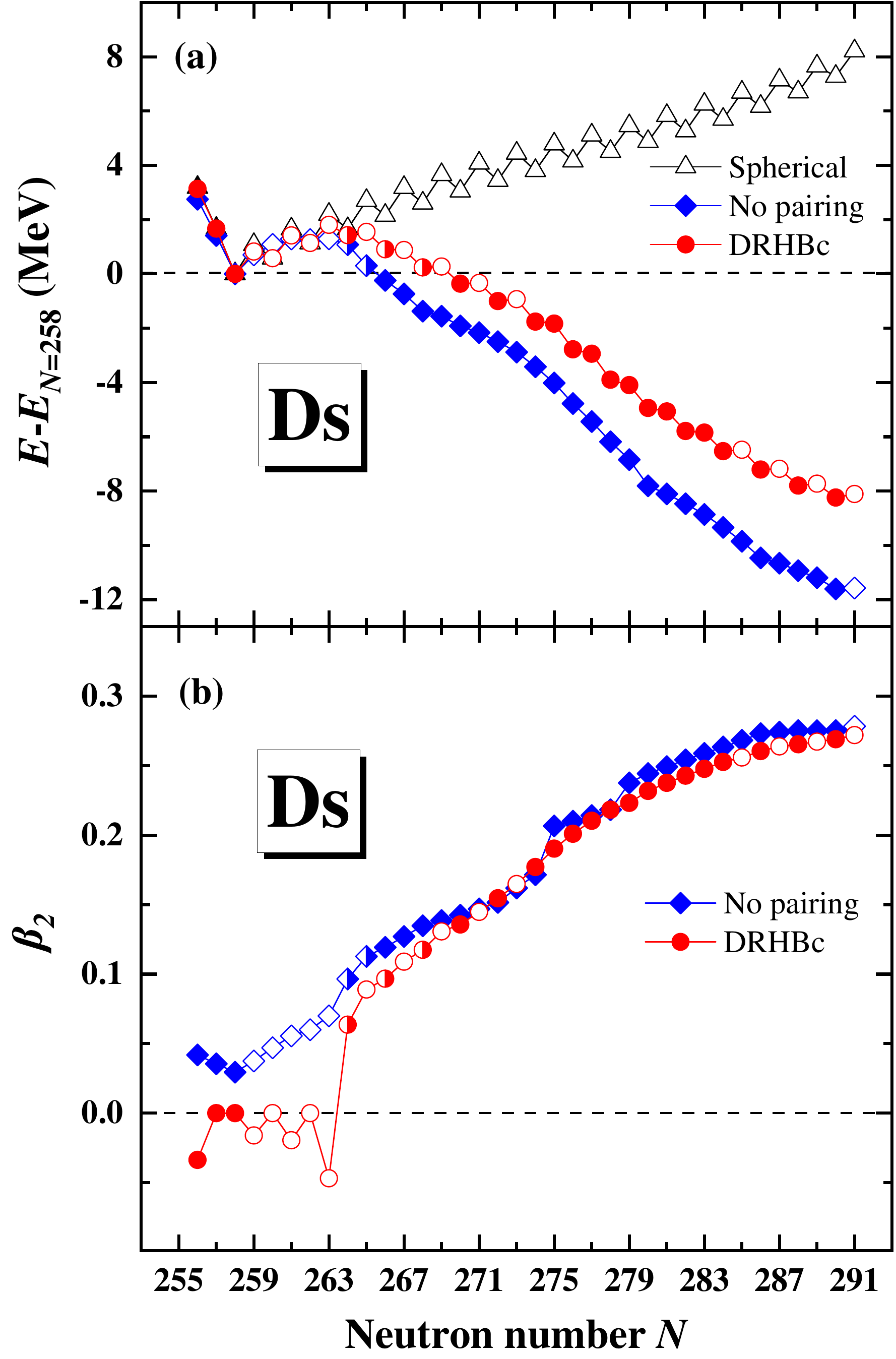}
	\caption{The DRHBc calculated (a) total energies of Ds isotopes near the drip line relative to that of $^{368}$Ds ($N = 258$) and (b) deformation parameters $\beta_2$ for these isotopes are shown by circles as a function of the neutron number $N$. The DRHBc results assuming spherical symmetry ($\lambda_{\mathrm{max}}=0$) and neglecting pairing correlations are respectively shown by triangles and diamonds. The solid, open, and semi-solid symbols represent nuclei that are bound, unbound, and bound against one-neutron emission but unbound against multi-neutron emission, respectively.}
	\label{fig3}
\end{figure}

We take Ds isotopes as examples to further study the odd-even difference in the one-neutron separation energy between odd-$N$ and even-$N$ nuclei.
Figure \ref{fig3} shows the DRHBc calculated total energies of Ds isotopes near the drip line relative to that of $^{368}$Ds$_{258}$ and their deformation parameters $\beta_2$.
The results from calculations assuming spherical symmetry and neglecting pairing correlations are also shown for comparison.
Here, the total energy equals negative binding energy, $E = -B$, namely, a lower total energy corresponds to a larger binding energy.
By adding neutrons, the decrease (increase) of the total energy means the stability (instability) against neutron emission of the obtained nucleus.
It can be found in Fig. \ref{fig3}(a) that the variation of the total energy versus $N$ shows clear odd-even staggering in the DRHBc results.
In particular, the total energy increases from an even-$N$ nucleus to the next odd-$N$ nucleus in regions of $258 \leqslant N \leqslant 273$ and $284 \leqslant N \leqslant 291$, indicating the instability against one-neutron emission of these odd-$N$ nuclei.
As a result, the number of bound odd-$N$ nuclei in the stability peninsula ($275 \leqslant N \leqslant 283$) is 5, much less than that of bound even-$N$ nuclei, 11.

In the DRHBc results shown in Fig. \ref{fig3}(b), the Ds isotopes with $N = 258, 260,$ and $262$ are spherical while those with $N =259,261,$ and $263$ exhibit slight oblate deformation with small negative $\beta_2$ due to the polarization effect.
For $N\geqslant 264$, $\beta_2$ becomes larger than $0.05$ and continues to increase with $N$, thereby neutron emitters and bound nuclei emerging.
In the spherical results shown in Fig. \ref{fig3}(a), the total energy of $^{368}$Ds is lower than those of heavier isotopes, which means that without deformation no bound nucleus exists beyond the neutron drip line.
Therefore, the deformation effects, which make the nuclear system more stable by lowering its total energy, play an indispensable role in the appearance of stability peninsula.
Meanwhile, the spherical results exhibit more pronounced odd-even differences in the total energy compared with the deformed results.

In the results without pairing shown in Fig. \ref{fig3}(a), after $N = 263$ the total energy keeps decreasing gradually until $N = 291$.
No visible odd-even difference is found in the total energy, and the numbers of bound odd-$N$ and even-$N$ nuclei only differ by 1.
Therefore, the odd-even differences between odd-$N$ and even-$N$ nuclei stem from the effects of pairing correlations.
Both deformation and pairing effects contribute to the sophisticated phenomenon of reentrant stability with odd-even differences.
It can also be found in Fig. \ref{fig3}(b) that without pairing most nuclei become more deformed, indicating that pairing correlations might reduce nuclear deformation, which is consistent with the DRHBc + PC-PK1 results for the halo nucleus $^{19}$B \cite{Sun2021PRC(1)}.

\begin{figure}[htbp]
	\includegraphics[scale=0.37]{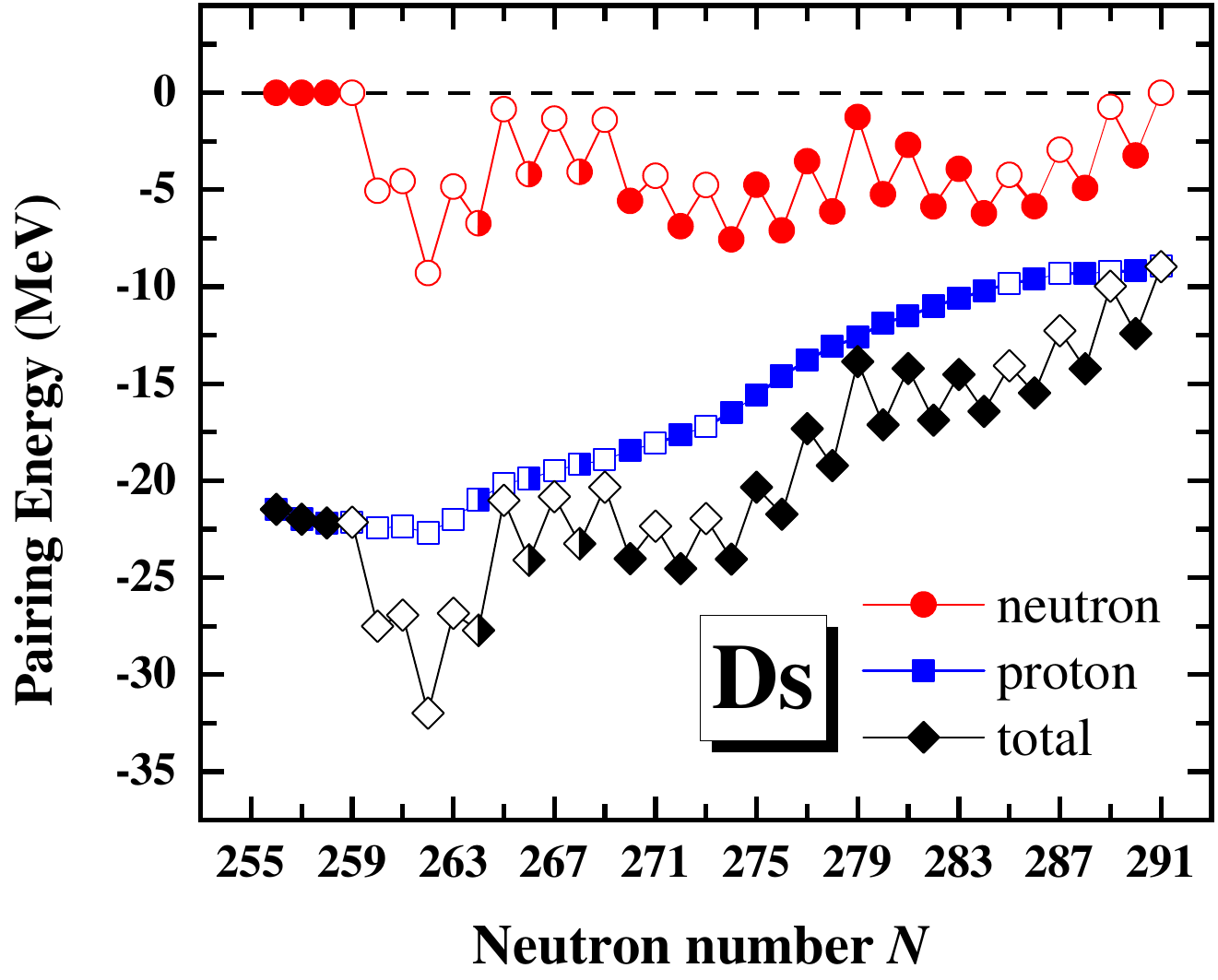}
	\caption{The DRHBc calculated neutron, proton, and total pairing energies as functions of the neutron number $N$ for Ds isotopes near the drip line. The solid, open, and semi-solid symbols represent nuclei that are bound, unbound, and bound against one-neutron emission but unbound against multi-neutron emission, respectively.}
	\label{fig4}
\end{figure}

To further examine the contribution of pairing, we show in Fig. \ref{fig4} the neutron, proton, and total pairing energies of Ds isotopes near the drip line.
While the proton pairing energy varies steadily with increasing $N$, there are obvious odd-even differences in the neutron pairing energy.
Due to the blocking effects of the unpaired neutron, the neutron pairing energy of an odd-$N$ nucleus is smaller than those of its even-$N$ neighbors in absolute value.
This results in odd-even differences in the total pairing energy, the total energy, the one-neutron separation energy, and finally the stability.

\begin{figure*}[htbp]
	\includegraphics[width=1\textwidth]{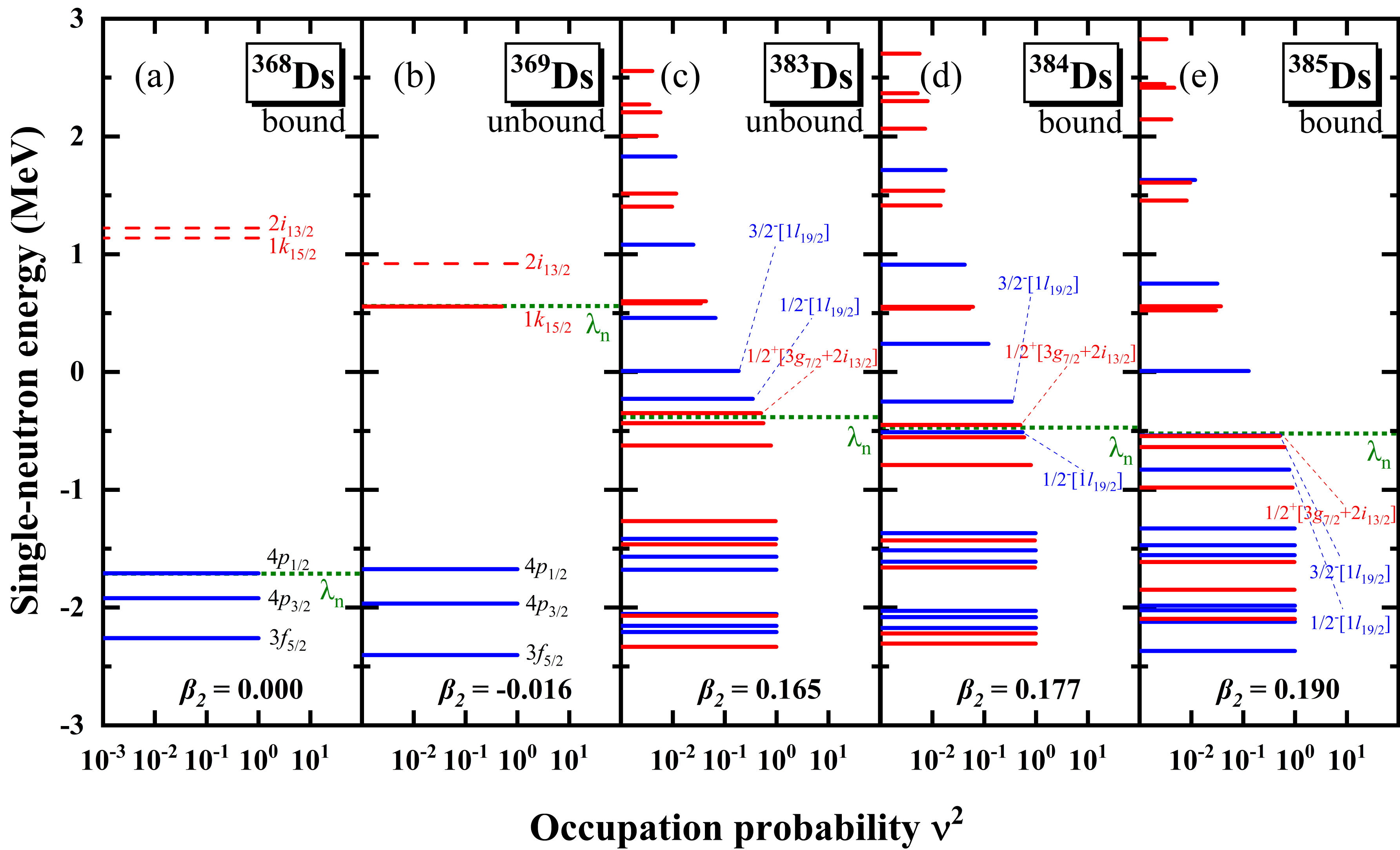}
	\caption{Single-neutron orbitals in the canonical basis around the Fermi energy $\lambda_n$ (dotted line) versus the occupation probability $\nu^2$ for $^{368,369,383,384,385}$Ds in the DRHBc calculations. For the spherical (near-spherical) nucleus $^{368}$Ds ($^{369}$Ds), orbitals are labeled by their (approximate) quantum numbers $nlj$, and the unoccupied orbitals are shown by dashed lines. For deformed nuclei $^{383,384,385}$Ds, orbitals around $\lambda_n$ are labeled by quantum numbers $\Omega^\pi$, and their major components are given in the square brackets.}
	\label{fig5}
\end{figure*}

To investigate the microscopic mechanism behind the reentrant stability, we illustrate the single-neutron orbitals around the Fermi energy for $^{368,369,383,384,385}$Ds in Fig. \ref{fig5}.
In Fig. \ref{fig5}(a), the gap between the highest occupied orbital and the lowest unoccupied orbital in the spherical nucleus $^{368}$Ds is approximately 3 MeV, supporting a shell closure at $N = 258$.
The valence neutrons in this nucleus occupy the bound orbital $4p_{1/2}$.
With one neutron added, the nucleus $^{369}$Ds becomes slightly deformed, and its valence neutron occupies an orbital split from $1k_{15/2}$ that is embedded in continuum, making it unstable against one-neutron emission.
With more neutrons added, the nuclei deviate more from spherical shape.
In Fig. \ref{fig5}(c), the nucleus $^{383}$Ds is deformed with $\beta_2 = 0.165$, and several down-sloping orbitals cross the continuum threshold and become bound.
Due to the loss of the pairing energy as shown in Fig. \ref{fig4}, the total energy of $^{383}$Ds is higher than that of $^{382}$Ds.
In Fig. \ref{fig5}(c), with one more neutron added, the deformation parameter for $^{384}$Ds increases to 0.177, and one more negative-parity orbital, $3/2^-$, becomes bound.
The one-neutron, two-neutron, and multi-neutron separation energies of $^{384}$Ds are all positive, making it a bound nucleus.
In Fig. \ref{fig5}(d), the deformation parameter increases further to 0.190 for $^{385}$Ds.
Although the pairing energy is suppressed by the blocking effects, the extra total energy lowered by deformation makes $^{385}$Ds stable against one-neutron emission.
Therefore, while nuclear deformation plays an essential role in the reentrant stability by significantly affecting the shell structure, the interplay between deformation and pairing effects influences the position where the odd-$N$ nucleus becomes bound in this peninsula.

\begin{figure*}[htbp]
	\includegraphics[width=0.95\textwidth]{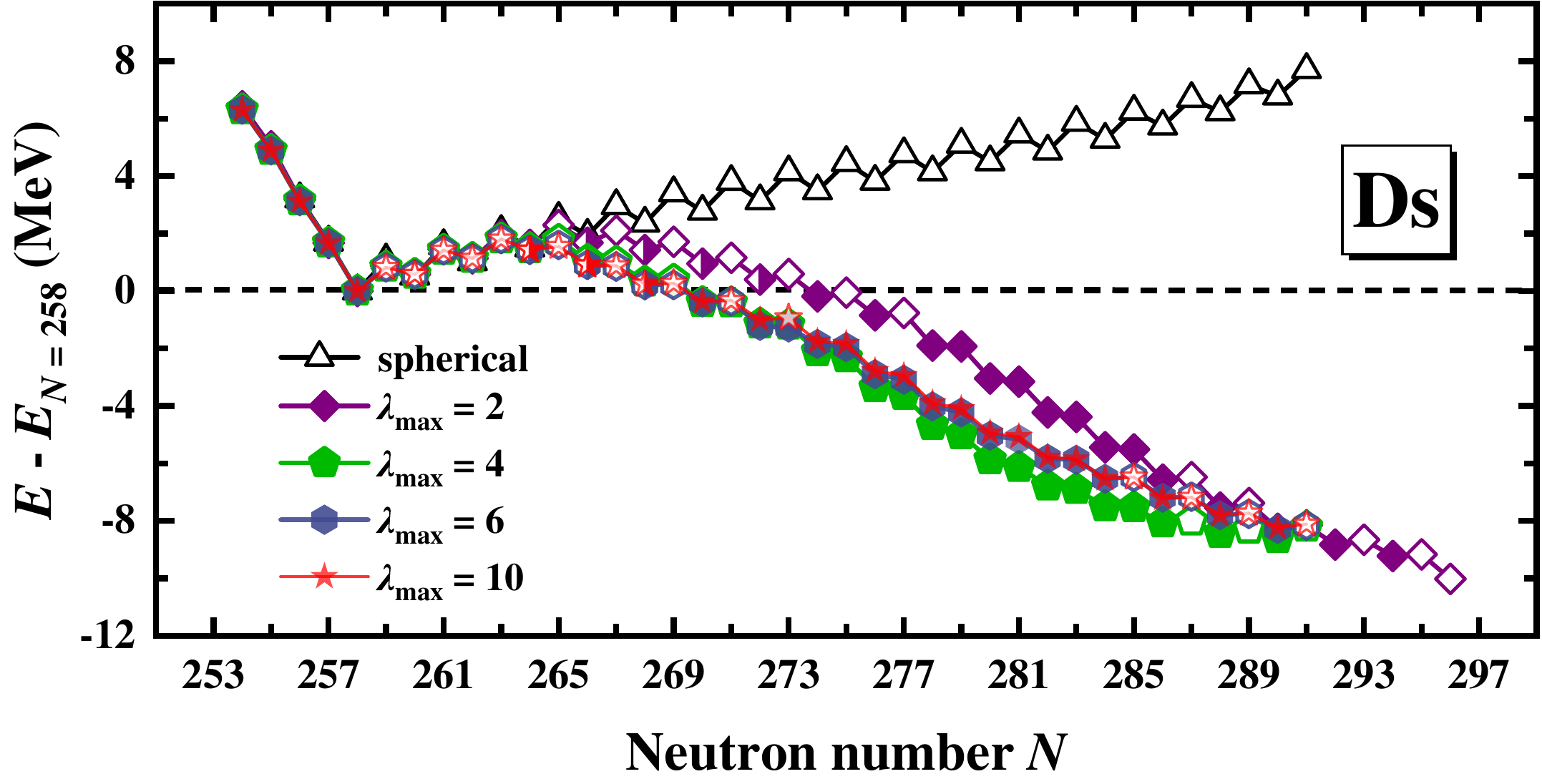}
	\caption{The DRHBc calculated total energies of Ds isotopes near the drip line relative to that of $^{368}$Ds as functions of the neutron number $N$ with the Legendre expansion \eqref{Legendre} truncated at different orders. The solid, open, and semi-solid symbols represent nuclei that are bound, unbound, and bound against one-neutron emission but unbound against multi-neutron emission, respectively.}
	\label{fig6}
\end{figure*}

It has been demonstrated that the effects of higher-order deformations are crucial for the ground-state properties of superheavy nuclei in a previous CDFT study \cite{Wang2022CPC}.
To reveal the effects of deformation at different order, we show in Fig. \ref{fig6} the calculated total energies of Ds isotopes near the drip line relative to that of $^{368}$Ds with the Legendre expansion \eqref{Legendre} truncated at different orders.
Namely, $\lambda_{\mathrm{max}}$ = 2 corresponds to the calculation only including quadrupole deformation $\beta_2$, $\lambda_{\mathrm{max}}$ = 4 corresponds to the calculation including quadrupole and hexadecapole deformations $\beta_2$ and $\beta_4$, and so on.
As already seen in Fig. \ref{fig3}, when assuming spherical symmetry, there is no bound nucleus beyond $N=258$ but clear odd-even difference in the total energy.
After including quadrupole deformation, the reentrant stability appears, with 11 even-$N$ and 4 odd-$N$ bound nuclei.
It can be also found that the odd-even differences in the total energy are reduced.
After further including hexadecapole deformation, the odd-even staggering in total energy appear further reduced, and the number of bound odd-$N$ nuclei in the stability peninsula increases to 8.
While the number of even-$N$ nuclei keeps unchanged as 11 with increasing $\lambda_{\mathrm{max}}$, that of bound odd-$N$ nuclei decreases respectively to 6 and 5 in the results with $\lambda_{\mathrm{max}} = 6$ and 10.
Actually, the results with $\lambda_{\mathrm{max}} = 6$ and 10 look quite close, indicating that the effects of $\beta_8$ and $\beta_{10}$ are marginal for the stability peninsula.
In other words, from a view of expansion, the results suggest that the calculations are essentially converged with $\lambda_{\mathrm{max}}$.

Finally, we note that the DRHBc calculations also predict many neutron emitters in the peninsula of stability. The neutron radioactivity is one of the most appealing topics in modern nuclear physics \cite{Pfutzner2012RMP}.
For instance, the two-neutron emitter $^{16}$Be observed in 2012 \cite{Spyrou2012PRL} has been further investigated recently \cite{Monteagudo2024PRL}.
Another recent experiment that discoveries new isotopes $^{160}$Os and $^{156}$W has revealed enhanced stability of the $N = 82$ shell closure toward the proton drip line by analyzing $\alpha$-decay energies and half-lives \cite{Yang2024PRL}. This finding implies that the unobserved $^{164}$Pb, despite being beyond the proton drip line, could be a doubly magic nucleus with increased stability \cite{Yang2024PRL}. 
The half-life estimation can be achieved by the DRHBc + WKB approach, which has been successful in describing $\alpha$-decay half-lives of even-even nuclei with $74\leqslant Z\leqslant 92$ \cite{Choi2023arXiv} and the one-proton emission half-life of $^{149}$Lu \cite{Xiao2023PLB} as well as predicting the multi-neutron emission half-lives of bound Ba and Sm isotopes beyond the neutron drip line \cite{Pan2021PRC}. Therefore, it is quite encouraging to estimate the half-lives of the DRHBc predicted neutron emitters in future works, which may provide reference for experimental search of possible stability beyond the drip line.

\section{Summary}\label{summary}

In summary, the predictive power of the DRHBc theory is demonstrated for superheavy nuclear masses in the region of $101 \leqslant Z \leqslant 118$.
For the 36 (240) measured (measured plus evaluated) mass data compiled in AME2020, the DRHBc description accuracy is 0.763 (0.833) MeV, in contrast to 0.339 (1.302) MeV by WS4 and 0.692 (2.809) MeV by FRDM.
The DRHBc theory predicts 93 bound nuclei with $106 \leqslant Z \leqslant 112$ beyond the primary neutron drip line $N = 258$, which form a stability peninsula adjacent to the nuclear landscape.
Significant odd-even differences between odd-$N$ and even-$N$ nuclei are found in the peninsula;
an odd-$N$ nucleus is less stable than its neighboring even-$N$ nuclei with a smaller one-neutron separation energy.
As a result, the reentrant stability for odd-$N$ nuclei appears later and ends earlier than that for even-$N$ nuclei along an isotopic chain.

The underlying mechanism behind the formation of stability peninsula and the odd-even difference are investigated in detail by taking Ds isotopes near the drip line as examples.
There is no reentrant stability in the results from calculation assuming spherical symmetry, demonstrating that the deformation effects play an indispensable role in the formation of stability peninsula.
In the results without pairing, the total energy varies smoothly between odd-$N$ and even-$N$ nuclei, demonstrating that the odd-even differences stem from the effects of pairing.
Specifically, the neutron pairing energy in an odd-$N$ nucleus is suppressed by the blocking effects of the unpaired neutron.

By investigating the single-neutron orbitals around the Fermi energy for $^{368,369,383,384,385}$Ds, it is found that the deformation effect significantly affects the nuclear shell structure.
With the increasing deformation, more and more down-sloping orbitals cross the continuum threshold and become bound.
The even-$N$ nucleus $^{380}$Ds first become bound due to this deformation effect, but its neighboring odd-$N$ nuclei are still unbound.
With more neutrons added, the deformation increases further, and the odd-$N$ nucleus $^{380}$Ds gains enough binding energy from the deformation effect to become stable against one-neutron emission.
The results suggest that the interplay between deformation and pairing effects can influence the position where the odd-$N$ nucleus becomes bound in the stability peninsula.

By examining the deformation effect at different orders, it is found that quadrupole deformation $\beta_2$ makes the most important contribution to the reentrant stability beyond the neutron drip line.
Hexadecapole deformation $\beta_4$ weakens the odd-even differences and, thus, increases the number of odd-$N$ nuclei in the peninsula.
The results considering deformations up to $\beta_6$ and to $\beta_{10}$ appear very close, indicating that the effects of $\beta_8$ and $\beta_{10}$ are marginal for the stability peninsula and the calculations are essentially converged with $\lambda_{\mathrm{max}}$ from a view of expansion.

\section*{Acknowledgement}

Helpful discussions with members of DRHBc Mass Table Collaboration are gratefully acknowledged. This work is supported by the National Natural Science Foundation of China (Grants No. U2032138, No. 11775112 and No. 12305125), the National Key R$\&$D Program of China (Contract No. 2023YFA1606503), the Natural Science Foundation of Sichuan Province (Grant No. 24NSFSC5910), and the High-performance Computing Platform of Huzhou University.

\bibliography{ref}

\end{document}